\documentclass[epj]{webofc}
\usepackage[varg]{txfonts}   
\newcommand{\beq}{\begin{equation}}
\newcommand{\eeq}{\end{equation}}
\newcommand{\beqa}{\begin{eqnarray}}
\newcommand{\eeqa}{\end{eqnarray}}
\DeclareMathOperator{\im}{Im}

\woctitle{Resonance Workshop in Catania}
\begin{document}
\title{Vector Meson Spectral Functions in a Coarse-Graining Approach}

\author{Stephan Endres\inst{1,2}\thanks{\email{endres@fias.uni-frankfurt.de}} \and
        Hendrik van Hees\inst{1,2} \and
        Janus Weil\inst{1,2} \and
        Marcus Bleicher\inst{1,2}
}

\institute{Institut für Theor. Physik, Universität Frankfurt,
Max-von-Laue-Straße 1, D-60438 Frankfurt, Germany
\and
Frankfurt Institute for Advanced Studies,
Ruth-Moufang-Straße 1, D-60438 Frankfurt, Germany
          }

\abstract{%
Dilepton production in heavy-ion collisions at top SPS energy is investigated within a 
coarse-graining approach that combines an underlying microscopic evolution of the nuclear 
reaction with the application of medium-modified spectral functions. Extracting local 
energy and baryon density for a grid of small space-time cells and going to each cell's 
rest frame enables to determine local temperature and chemical potential by application of an equation 
of state. This allows for the calculation of thermal dilepton emission. We apply and compare two 
different spectral functions for the $\rho$: A hadronic many-body calculation and 
an approach that uses empirical scattering amplitudes. Quantitatively good 
agreement of the model calculations with the data from the NA60 collaboration is achieved 
for both spectral functions, but in detail the hadronic many-body
approach leads to a better 
description, especially of the broadening around the pole mass of the $\rho$ 
and for the low-mass excess. We further show that the presence of a pion 
chemical potential significantly influences the dilepton yield.
 }
\maketitle
\section{Introduction}
\label{intro}
 
Although lepton pairs came into the focus of interest as probes for the
creation of a deconfined phase
\cite{Shuryak:1978ij,Domokos:1980ba,McLerran:1984ay}, the so-called
quark-gluon plasma (QGP), the hadronic contributions to dilepton
production have become a field of vigorous theoretical investigation on
their own \cite{Hatsuda:1991ez, Klingl:1997kf, Rapp:1999ej,
  Leupold:2009kz}. As vector mesons directly couple to the
electromagnetic current and can convert into a lepton pair, the change
of their spectral shape inside hot and dense matter can be studied by
measuring invariant-mass spectra of dileptons. These properties of
hadrons in the medium are of immanent interest to achieve a full
understanding of the phase structure given by quantum chromodynamics
(QCD). Detailed investigations might help to understand how the 
high-energy regime with partonic degrees of freedom is connected to the
non-perturbative low-energy part, where quarks and gluons are confined in
hadrons. Another aspect of study is the symmetry pattern of QCD which is
expected to change from its vacuum properties when going to finite temperature and
baryochemical potential. In particular, the restoration of chiral symmetry
(which is spontaneously broken in the vacuum) is predicted. However, a
theoretical and experimental clarification of these issues is associated
with several theoretical challenges \cite{Rapp:1999ej,
  Leupold:2009kz,Gale:1987ki}:
\begin{itemize}
\item Due to the fact that leptons do not interact strongly, dilepton
spectra always represent an integral over the whole evolution of the
collision. In the course of the reaction, dileptons can be produced in
many different ways. Therefore one has to identify the relevant sources
and disentangle a large number of contributions, which is by no means
trivial.
\item Consequently, the dilepton production is also highly influenced by
the reaction dynamics. This concerns the lifetime and expansion of the
created fireball but also the spatial distribution of temperature and
density. Furthermore, the question if and how fast the system will
equilibrate is of importance.
\item As the reactions in a heavy-ion collision are governed by the
strong interaction, one has to deal with complications such as
multi-particle collisions, finite temperature and density corrections to
form factors and widths, off-shell dynamics, etc.
\end{itemize} 
These aspects impose high requirements on theoretical descriptions that
model the dilepton production in heavy-ion collisions. It is necessary
(i) to identify the relevant contributions to the dilepton yield, (ii)
find a realistic description of the space-time evolution of the reaction
and (iii) implement the medium modifications in an appropriate way.

While the dilepton sources at SPS energies are more or less
uncontentious, there exist several different approaches providing a
theoretical description of the reaction dynamics. In fireball models
\cite{vanHees:2006ng,vanHees:2007th} the zone of hot and dense matter is
described by an isentropically expanding cylindrical volume. Here an
implementation of medium effects is straightforward via the application
of spectral functions from thermal quantum field theory. But such a
fireball parametrization gives a simplified picture of the reaction,
with one value for temperature $T$ and baryochemical potential $\mu_{B}$
in the whole volume. A more nuanced description is delivered by
hydrodynamic models \cite{Vujanovic:2013jpa,Ryblewski:2015hea}
respectively a hydro+transport hybrid model \cite{Santini:2011zw}. The
difficulty here is that one usually needs some initial state for the
start of the hydrodynamic evolution and additionally some description of
the final-state interactions. Furthermore, it is questionable whether
such approaches will also work at low energies, e.g. for SIS 18
energies.

A completely different ansatz is used in transport models
\cite{Thomere:2007cj,Bratkovskaya:2008bf,Linnyk:2011hz,Bratkovskaya:2013vx,Weil:2012ji,Vogel:2007yu,Schmidt:2008hm}. 
These approaches treat the dynamics microscopically and account for
non-equilibrium, but the implementation of full medium effects, which
were discussed above (i.e.\,multi-particle collisions, off-shell
effects, etc.), is difficult here.  Nevertheless, there have been some
successful investigations with medium-effects implemented in transport
models \cite{Cassing:1997jz,Bratkovskaya:1997mp,Linnyk:2011hz,Schenke:2005ry}. However, 
problems such as the large variety of parameters, especially cross-sections 
and branching ratios for which little data is available, remain. 

The idea of the present work is to combine a realistic microscopic
description of the reaction with an application of medium-modified
spectral functions\footnote{As we extract only baryon, pion and energy density for each space-time cell, we are not that sensitive with regard to all the details of the model parameters (as for example branching ratios) as an explicit transport-only calculation would be. Furthermore, we avoid the difficulties which a full off-shell description of the reaction would imply.}. Ideally, it is applicable at all stages of
the reaction so that one has a unified description well-working for
different collision energies (i.e. from SIS to RHIC or LHC). This led
to the development of the coarse-graining approach as presented in these
proceedings, following an ansatz proposed in
\cite{Huovinen:2002im}. 

This paper is structured as follows: Section \ref{Approach} introduces
the coarse-graining approach while section \ref{SF} gives an overview of
the spectral functions, discussing the various dilepton sources that
enter the calculation. An overview of the results for top SPS energy and
a comparison with the experimental dilepton spectra from the NA60
Collaboration is presented in section \ref{Results}. Finally, a summary
and outlook is given in section \ref{Outlook}.
\section{\label{Approach} The Coarse-Graining Approach} 

To combine a realistic microscopic (3+1)-dimensional description of the
colliding system with full in-medium spectral functions, one first has
to extract the local thermodynamic properties from the underlying
simulations.  For the present study, we use input from the
Ultra-relativistic Quantum Molecular Dynamics (UrQMD) approach
\cite{Bass:1998ca,Bleicher:1999xi,Petersen:2008kb}. It is a
non-equilibrium model which includes baryonic and mesonic degrees of
freedom with masses up to 2 GeV/$c^{2}$. The hadrons are propagated on
classical trajectories, and the particle production is described by
inelastic cross sections. String excitation is possible for collisions
with $\sqrt{s} > 3$\,GeV.

To obtain the thermodynamic properties from UrQMD calculations, it is
necessary to average over a sufficiently large number of events. For the
present study we use an ensemble of 1000 UrQMD events. Then one finds a
(relatively) smooth form of the particle distribution function,
\begin{equation} f(\vec{x},\vec{p},t)=\left\langle \sum_{h}
\delta^{3}(\vec{x}-\vec{x}_{h}(t))\,\delta^{3}(\vec{p}-\vec{p}_{h}(t))\right\rangle.
\end{equation} 
by summing over the phase-space contributions of each hadron $h$ and
taking the ensemble average. As the underlying dynamics is due to a
non-equilibrium model, it is only locally and only approximately
possible to extract the equilibrium properties. In practice we obtain
local energy and baryon density by setting up a space-time grid of small
cells\footnote{Note that with the size of the cells and the number of
  events chosen for the calculation, we get an approximate error in
  baryon density of roughly 10\% for $\rho=0.5\rho_{0}$, and of the same
  order also for energy density $\varepsilon$. This guarantees a
  sufficient accuracy of the simulations. In cells with higher
  densities, which are the dominant sources of thermal dileptons, one
  will find even less fluctuations. The error especially in temperature
  will be quite small, as $T \propto \varepsilon^{1/4}$. Furthermore,
  any fluctuations will be averaged out in the final (space-time
  integrated) dilepton spectra so that the overall statistical effect on
  the thermal yields is rather negligible. To obtain sufficient
  statistics for the non-thermal $\rho$ contribution (see below) several
  coarse-graining runs with different ensembles of UrQMD events as input
  had to be performed.} with dimensions
$\Delta x=\Delta y=\Delta z=0.8$\,fm and a time-step with
$\Delta t=0.2$\,fm/c.  For each cell, the averaged baryon four-flow
$j^{\mu}_{\mathrm{B}}$ and the energy-momentum tensor $T^{\mu\nu}$ can
be determined as
\begin{eqnarray} \ \ \ \ \ T^{\mu\nu}&=&\int
\mathrm{d}^{3}p\frac{p_{\mu}p_{\nu}}{p_{0}}f(\vec{x},\vec{p},t)
=\frac{1}{\Delta V}\left\langle \sum\limits_{i=1}^{N_{h} \in \Delta V}
\frac{p_{\mu}^{i}\cdot p_{\nu}^{i}}{p_{0}^{i}}\right\rangle, \\ \ \ \ \
\ j_{\mu}^{\text{B}} &=& \int \mathrm{d}^{3}p\frac{p_{\mu}}{p_{0}}
f^{\text{B}}(\vec{x},\vec{p},t) =\frac{1}{\Delta V}\left\langle
\sum\limits_{i=1}^{N_{\text{B}/\bar{\text{B}}} \in \Delta
V}\pm\frac{p_{\mu}^{i}}{p_{0}^{i}}\right\rangle.
\end{eqnarray} 
According to Eckart's definition, we can then find the local rest frame
of the cell by performing a Lorentz boost to a frame, where
$\vec{j}_{\mathrm{B}}=0$. The energy-momentum tensor should then be
diagonal and we can identify the energy density $\varepsilon$ with
$T^{00}$ while the baryon density is given by $j^{0}_{\mathrm{B}}$. The
assumption of an isotropic equilibrated thermal system can be checked by
looking at the diagonal matrix elements $T^{ii}$ ($i \in \{1,2,3\}$) of
the energy-momentum tensor which represent the transverse and
longitudinal pressures. While after a certain time they are usually
found to be roughly equal, at the very beginning of the collision (for
top SPS energy the first 1-2\,fm/$c^{2}$) one observes large pressure
differences. It is clear that when the two nuclei start to interact and
traverse each other at first a strong longitudinal pressure builds up
before the system becomes more or less isotropic at later times. To
account for this, we apply a description developed for anisotropic
hydrodynamics \cite{Florkowski:2012pf} which translates the anisotropic
momentum distribution into an equilibrium picture and gives a realistic
value of the "effective" energy density in the cell. Applying this
procedure allows to use the coarse-graining approach over the whole
evolution of the collision.

Once energy and baryon density in the cell are known, we need an
equation of state (EoS) to determine temperature and baryochemical
potential. For consistency with the underlying microscopic dynamics we
employ a hadron-gas EoS \cite{Zschiesche:2002zr} which includes the same
degrees of freedom as the UrQMD model. However, as we consider a
collision energy where the expected temperatures will be high enough to
create a deconfined phase of quarks and gluons, a pure hadron gas will
not give the correct description for high energy densities. Therefore we
use a second EoS obtained from a lattice fit \cite{He:2011zx} to take
the (cross-over) phase transition into account. The critical temperature
$T_{c}$ is here found to be at 170 MeV. We match the values of $T$ from
the two EoS in the temperature range from 150 to 170 MeV and apply the
lattice EoS exclusively for higher temperatures to avoid discontinuities
in the evolution. Note that in the lattice EoS $\mu_{B}$ is always zero.

Finally we calculate the thermal dilepton emission rate for each
cell. The dilepton yield per four-volume and four-momentum is directly
related to the imaginary part of the retarded electromagnetic
current-current correlator as \cite{McLerran:1984ay}
\begin{equation} \frac{\mathrm{d} N_{ll}}{\mathrm{d}^4x\mathrm{d}^4q} =
-\frac{\alpha_\mathrm{em}^2 L(M)}{\pi^3 M^2} \; f^{\text{B}}(q \cdot
U;T) \; \im \Pi^{(\text{ret})}_\mathrm{em}(M, \vec{q};\mu_B,T).
\label{rate}
\end{equation} 
In the hadronic domain, the correlator $\Pi^{(\text{ret})}_\mathrm{em}$
is proportional to the vector-meson fields respectively their spectral
functions under the hypothesis that vector-meson dominance is valid. For
the present approach we calculate thermal emission for all cells with a
temperature above 50 MeV. For lower temperatures, where the densities
are so low that the assumption of a thermal emission becomes
questionable and the tabulated EoS we use is less accurate, we directly
take the $\rho$ mesons from the UrQMD transport calculation and let them
decay to dileptons. As we only calculate either thermal or non-thermal
contribution for each cell, we avoid double-counting.

It is important to note that equation (\ref{rate}) is only valid for the
case that the system is in chemical equilibrium with regard to the pion
density and no pion chemical potential builds up.  However, it has been
discussed whether this is really an appropriate description. Some
authors have argued that an over-dense pionic system is created in the
initial stage of the collision and that the fireball remains out of
equilibrium due to the long relaxation time
\cite{Kataja:1990tp,Bebie:1991ij}. In other models $\mu_{\pi}$ becomes
finite after the number of pions is fixed at the chemical freeze-out but
the system further cools and extends
\cite{Kolb:2002ve,vanHees:2007th}. In any case, such a chemical
off-equilibrium requires an additional fugacity factor
$z^{n}_{\pi}=\exp(n\mu_{\pi}/T)$ in the foregoing equation, with $n=2$
in the case of the $\rho$ contribution and $n=3,4,5$ for multi-pion
interactions (see section \ref{SF}).  This is due to the fact that
equation (\ref{rate}) is independent of the hadronic initial and final
states as only chemical potentials of conserved charges are considered
for its derivation, for which $Q_{i} = Q_{f}$ is always
complied. However, since the pion number $N_{\pi}$ is not a conserved
quantity, this assumption is inappropriate in case of a finite pion
chemical potential which implies that generally
$N_{\pi,i} - N_{\pi,f} \neq 0$. Note that for the quark-gluon plasma
$\mu_{\mathrm{B}}=\mu_{\pi}=0$ is assumed in the present work. We will
investigate the influence of a pion chemical potential on the dilepton
rates in these proceedings. For this, we extract $\mu_{\pi}$ for each
cell in Boltzmann approximation according to
\begin{equation} \mu_{\pi}= T\cdot\ln \left(
\frac{2\pi^{2}n_{\pi}}{g_{\pi}Tm^{2}
\mathrm{K}_{2}(m/T)} \right),
\end{equation} 
where $\mathrm{K}_{2}$ denotes the Bessel function of the second kind,
$n_{\pi}$ the pion density in the cell and $g_{\pi}$ the degeneracy
factor (which is 3 for pions). 

A more detailed description of the coarse-graining approach and how it
is used to calculate thermal dilepton emission can be found in reference
\cite{Endres:2014zua}.
\section{\label{SF} In-medium Spectral Functions} 

Once the thermodynamic properties of the cell are known, it is important
to describe the dependency of the spectral functions on these
parameters. Note that the focus is on the $\rho$ here, as the $\omega$
and $\phi$ mesons have a much longer lifetime and only show small medium
modifications at SPS energies. As calculations from first principles in
the non-perturbative hadronic domain of QCD are difficult and a
measurement of the spectral shape is only possible for the vacuum (via
$e^{+}e^{-}$-collisions), one has to rely on hadronic models for this
purpose. There exist several approaches for spectral functions of vector
mesons, but only few of those consider both the influence of finite
temperature and finite density on the in-medium self-energy of the
$\rho$ meson.  In the present work we employed and compare the work by
Rapp and collaborators \cite{Rapp:1999us} who used hadronic many-body
theory to implement medium modifications (Rapp SF), and another more
model-independent approach by Eletsky et al. \cite{Eletsky:2001bb}
where empirical scattering amplitudes are used to determine the spectral
properties (Eletsky SF).

The Rapp SF includes three different contributions for the calculation
of the total self-energy: Polarizations of the pion cloud
($\Sigma_{\pi\pi}$), interactions with mesons, i.e.\,pions, kaons and
rhos ($\Sigma_{\rho M}$) and the resonant scattering with nucleons
respectively the most abundant baryonic resonances ($\Sigma_{\rho B}$)
\cite{Rapp:2000pe}. The calculation also takes anti-baryons into account
as the interaction with the $\rho$ is the same as for a baryon. For this
purpose the effective baryon density
$\rho_{\mathrm{eff}}=\rho_{N}+\rho_{\bar{N}}+0.5(\rho_{B^{*}}+\rho_{\bar{B}^{*}})$
instead of the baryochemical potential enters the Rapp SF as $\mu_{B}$
only accounts for the net baryon density. Accordingly, the in-medium
propagator takes the form
\begin{equation}
D_{\rho}(M,q;T)=\left[M^{2}-m^{0}_{\rho}-\Sigma_{\pi\pi}-\Sigma_{\rho
M}-\Sigma_{\rho B}\right]^{-1}.
\end{equation} 
In the approach by Eletsky et al.\ the spectral behavior is calculated
under the more simplified assumption of a non-interacting gas of pions
and nucleons. Consequently, only the $\rho-\pi$ and $\rho-N$ scattering
amplitudes enter the in-medium self-energy contribution. The idea here
is that the medium modification of a particle's spectral function can be
related to the forward scattering amplitude $f$ of this particle on the
constituents of the medium \cite{Eletsky:1998kk} via the optical
theorem. While the low-energy amplitudes are saturated with resonances
from the partial-wave analysis by Manley and Saleski
\cite{Manley:1992yb} plus a background contribution, a Regge fit is
performed for high-energy scatterings. For the $\rho$ meson scattering
from hadron $a$ in the heat bath follows a contribution to the in-medium
self energy of the form
\begin{equation}
\label{expand} \Sigma_{\rho a} (E,p)=-4\pi \int
\frac{\mathrm{d}^{3}k}{(2\pi)^{3}}n_{a}(\omega)\frac{\sqrt{s}}{\omega}f_{\rho
a}(s).
\end{equation} 
Note that equation (\ref{expand}) is a low-density expansion (similar to
a virial expansion) and therefore fully applicable only if the system is
dilute enough. However, in the hot and dense stage it is questionable
whether this assumption still holds. Additionally the optical analogy -
on which the above expression is based - requires the particle's
wavelength to be significantly smaller than the average distance between
the constituents of the medium and the scattering angle to be small. In
contrast to the approach by Rapp et al, the influence of kaons, excited
baryons and hyperons on the $\rho$ self energy is neglected
here. Furthermore, no anti-baryons are
included in the calculation.
\begin{figure} 
\includegraphics[width=0.47\columnwidth]{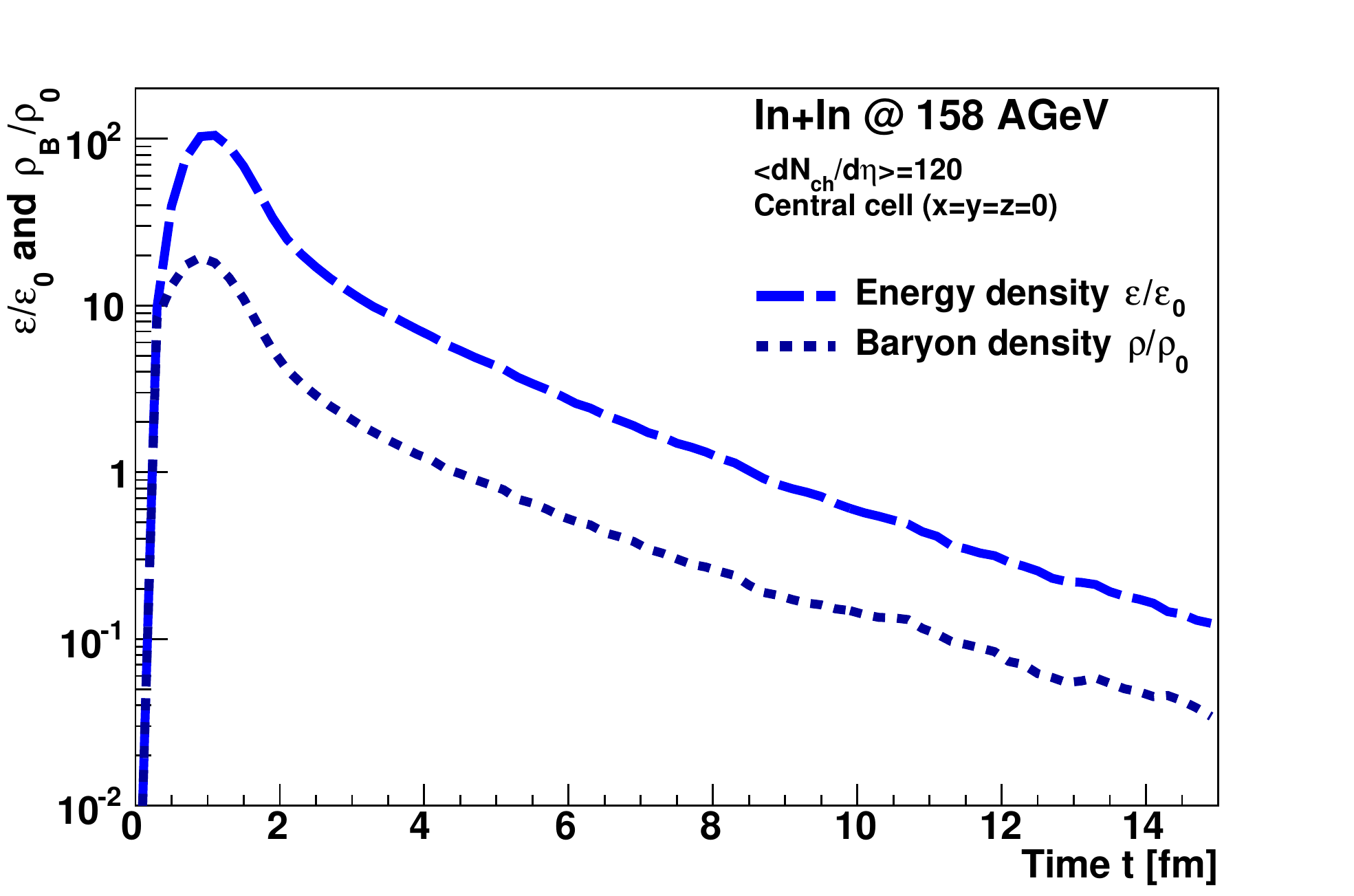} \hfill
\includegraphics[width=0.47\columnwidth]{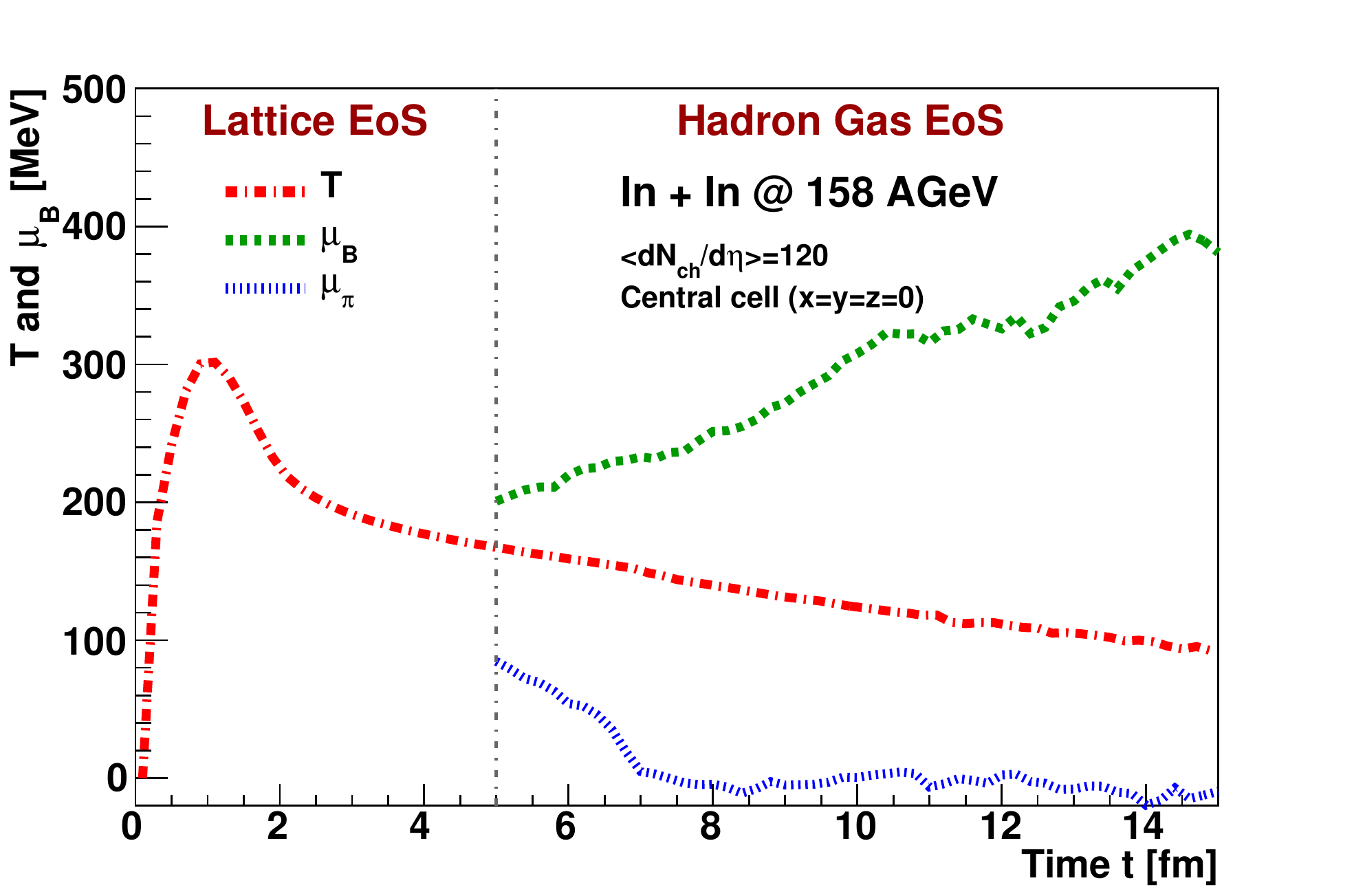}
\caption{Time evolution of the energy and baryon densities,
  $\varepsilon$ and $\rho_{B}$, in units of the ground-state densities
  (left) and for temperature $T$ as well as the baryon and pion chemical
  potentials $\mu_{B}$ and $\mu_{\pi}$ (right) for In+In collisions at
  158\,AGeV. The results shown here are from the central cell of the
  coarse-graining grid, i.e.\,for a small volume around $x=y=z=0$.}
\label{fig-1}
\end{figure}
Besides the $\rho$, it will be important to include also the other
relevant thermal sources (although we focus on the $\rho$ part here). At
the collision energy considered here, the creation of a deconfined phase
of free quarks and gluons is possible. In this case dilepton production
is possible via the reaction $q\bar{q} \rightarrow l^{+}l^{-}$, i.e.\, a
quark and an anti-quark can annihilate into a virtual photon
respectively a lepton pair. To determine the QGP dilepton yield we use
an extrapolation of rates obtained from lattice calculations
\cite{Ding:2010ga} to non-zero three momentum \cite{Rapp:2013nxa}. In
the hadronic phase, a significant contribution to the dilepton yield will
also originate from multi-pion interactions. Following the approach of
Dey, Eletsky and Ioffe \cite{Dey:1990ba} this contribution is determined
taking the mixing of the vector and axial-vector currents into account.
\section{\label{Results} Results} 

For the results presented here we used as input 1000 events of In+In
collisions at a beam energy of $E_{\mathrm{lab}}=158$\,AGeV simulated
with the most recent version 3.4 of the UrQMD model. We chose an impact
parameter distribution from 0 to 8.5\,fm, corresponding to a charged
particle number
$\left\langle \mathrm{d}N_{\mathrm{ch}}/\mathrm{d}\eta \right\rangle =
119$
which is very close to the experimentally measured value.

Figure \ref{fig-1} shows the time evolution for the central cell of the
coarse-graining grid, i.e.\,for a small volume around $x=y=z=0$. While
the left plot presents the energy and baryon densities in units of the
ground state densities, the right plot depicts the evolution of
temperature $T$ as well as the baryon and pion chemical potentials
$\mu_{B}$ and $\mu_{\pi}$. We obtain a maximum energy density that is
100 times the ground state density and a baryon density reaching up to
20 times $\rho_{0}$. The maxima are reached roughly 1\,fm/c after the
begin of the collision. Similarly, as is visible from the right plot, 
the temperature reaches its highest value of roughly 300 MeV after 
1\,fm/c and drops off again afterwards. However, even after 15\,fm/c the 
cell has still a temperature of 100 MeV.  However, one has to bear in mind 
that here in the center of the grid one obtains the locally highest densities 
which are not representative for the whole colliding system. Therefore this
behaviour is mainly a specific property of the central cell, more
peripherally the temperature is significantly lower. The dotted vertical
line around $t=5$\,fm/$c$ denotes the transition from the lattice EoS,
which is used for temperatures above 170 MeV, to the hadron gas. As we
assume $\mu_{\mathrm{B}}=\mu_{\mathrm{\pi}}=0$ in the QGP phase, we get
values for the chemical potentials only after the temperature
temperatures has fallen below $T_{c}$. While the baryon chemical
potential then slowly increases from a value of 200 MeV to 400 MeV we
observe an initially high pion chemical potential which drops to values
around zero in the course of the evolution. The latter effect is due to
the non-equilibrium nature of the underlying transport dynamics where
many pions are created at the beginning of the collision, especially via
string excitation. However, note once again that the central cell is not
necessarily representative for the evolution of the whole system.
\begin{figure} 
\includegraphics[width=0.47\columnwidth]{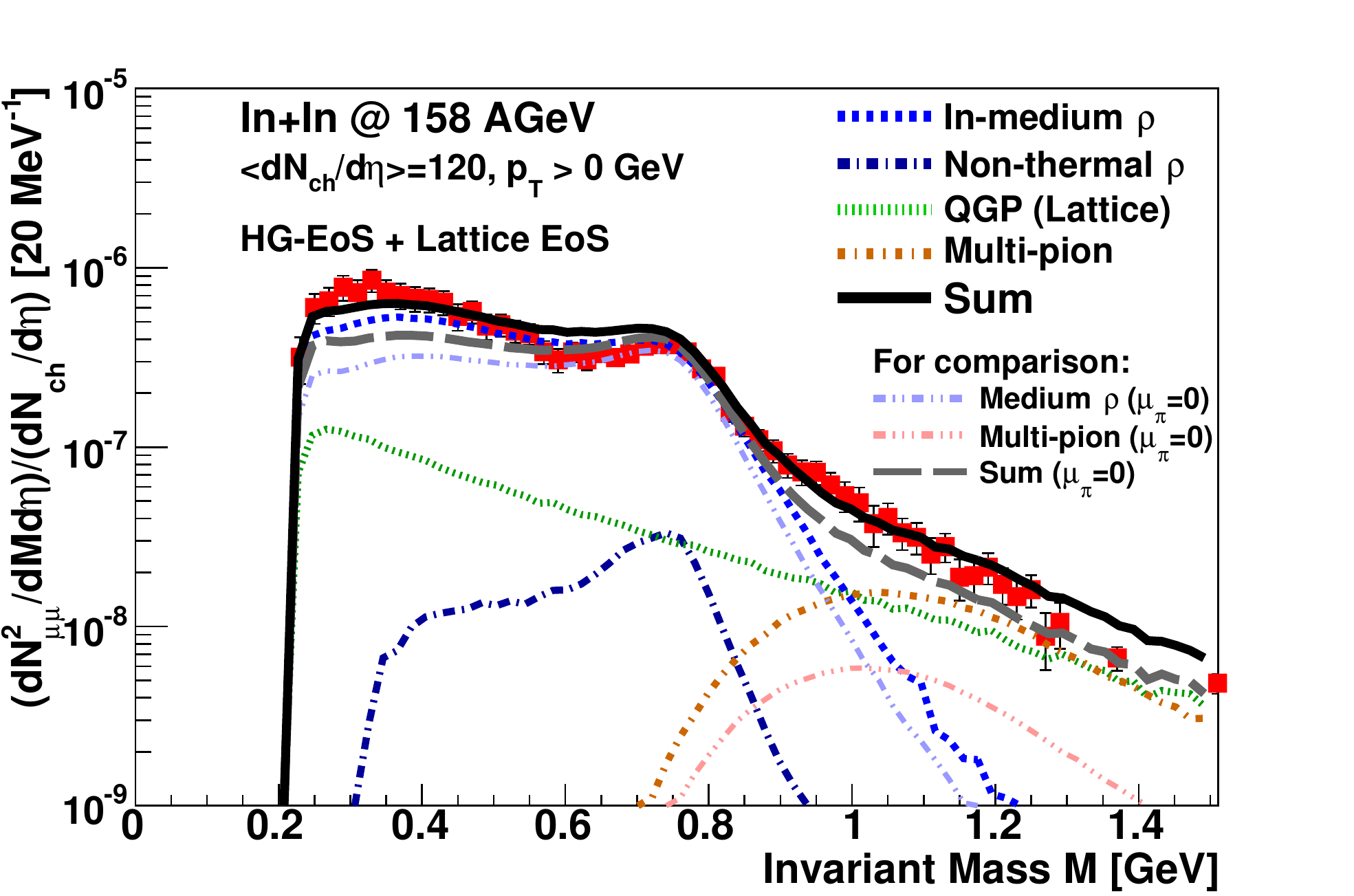}\hfill
\includegraphics[width=0.47\columnwidth]{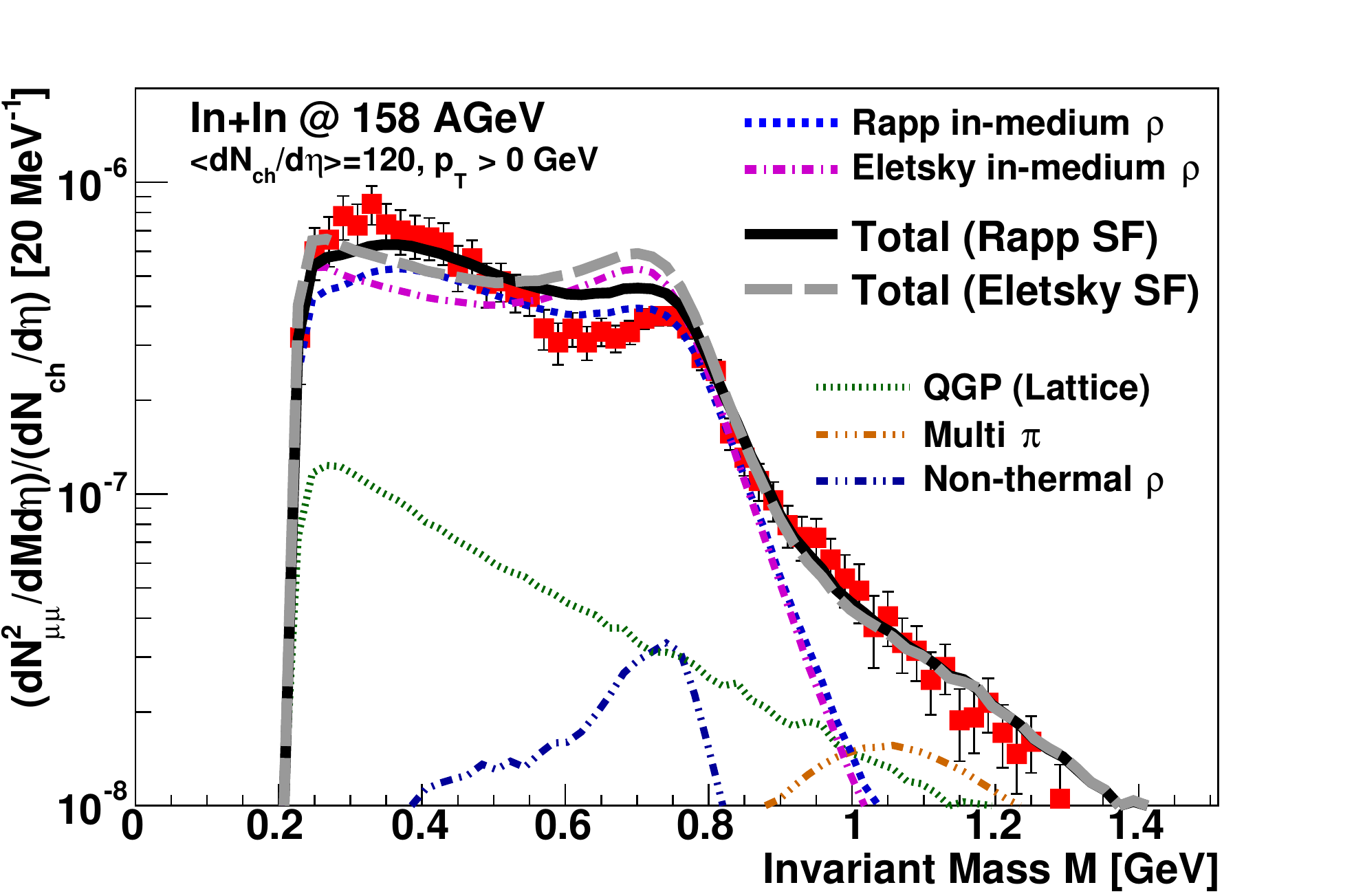}
\caption{(Left) Invariant-mass spectrum of the dilepton excess yield for
  In+In collisions at 158\,AGeV as obtained within the coarse-graining
  approach. We show the thermal $\rho$ contribution (Rapp SF) as well as
  the yield from multi-pion interactions and from the quark-gluon
  plasma. The model results are compared to the experimental data from
  the NA60 collaboration \cite{Arnaldi:2006jq}. For comparison, we also
  show the dilepton yield under the assumption of a vanishing pion
  chemical potential. (Right) Same as in the left plot, but here
  resulting yields from the two $\rho$ spectral functions by Rapp and
  Eletsky are compared to each other.}
\label{fig-2}
\end{figure}

The dilepton invariant mass spectra are presented in Figure
\ref{fig-2}. The left plot shows the result as obtained with the
coarse-graining approach and the Rapp spectral function for the
$\rho$. The total yield includes the thermal $\rho$ contribution as well
as the yield from multi-pion interactions, from the quark-gluon plasma
and a transport $\rho$ contribution from cold cells. The model results
are contrasted to the experimental data from the NA60 collaboration
\cite{Arnaldi:2006jq}. The description of the measured yield is quite
good within the error bars, especially the low-mass excess can be
explained with a broadening of the $\rho$ spectral function as compared
to the vacuum shape. The non-thermal $\rho$ is here rather negligible
in relation to the thermal contribution. Also the higher mass region above
1 GeV/$c^{2}$, which is dominated by the contributions from multi-pion
interactions and the QGP, can be described with the model. For
comparison, we also show the dilepton yield for the assumption of a
vanishing pion chemical potential. This results in an underestimation of
the dilepton yield in the region from 0.2 to 0.4 GeV/$c^{2}$ by a factor
of two. Note that this difference is primarily caused by the fugacity
factor in equation (\ref{rate}) which might easily double the emission
rate in the presence of a large pion chemical potential, whereas the
explicit dependence of the Rapp spectral function on $\mu_{\pi}$ is
rather moderate. The large effect on the low-mass tail is due to the
fact that the pion chemical potential reaches its highest values early
in the evolution, when the baryon density is still very high, which is
known to be the main cause of the low-mass enhancement
\cite{Rapp:1999ej,Leupold:2009kz,Endres:2014zua}. A significantly lower
yield from the multi-pion contribution is obtained as well for
$\mu_{\pi}=0$, so that in the mass range around 1\,GeV/$c^{2}$ the
theoretical prediction is clearly below the data then.

The comparison between the results with the two different approaches for
the $\rho$ spectral functions are shown in the right plot of Figure
\ref{fig-2}. While the peak at the pole mass of the $\rho$ meson is more
pronounced for the Eletsky SF, a stronger dilepton yield shows up in the
lower mass range from 0.3 to 0.5 GeV/c$^{2}$ when using the Rapp SF. In
general it accounts for a stronger melting of the peak due to baryonic
effects for the spectral function from hadronic many-body
theory. Naturally one will see a stronger density effect than within the
low-density expansion of the empirical scattering amplitudes, as used in
the approach of Eletsky. Besides, the Eletsky SF does not include all
the effects considered within the Rapp SF as only $\rho - N$ and
$\rho - \pi$ scatterings are considered. It has also been pointed out
that the Eletsky SF (where the amplitudes are evaluated on-shell only)
misses the off-shell contributions from sub-threshold baryon resonances,
that contribute substantially to the low-mass enhancement of the $\rho$
\cite{Rapp:1999ej}.
\section{\label{Outlook} Conclusions \& Outlook} 
We have presented calculations of the thermal dilepton yield for In+In
collisions at top SPS energy from coarse-grained microscopic
dynamics. The results show a good agreement with the measurements of the
NA60 collaboration. While both in-medium spectral functions for the
$\rho$ meson have proven to be in qualitative accordance with the data,
the Rapp SF shows a stronger broadening of the spectral shape and gives
a significantly better quantitative description. The present study also
shows that within the coarse-graining approach the assumption of a
finite pion chemical potential is crucial to describe the data. This is
in agreement with other investigations that included a pion chemical
potential \cite{Koch:1992rs,vanHees:2006ng}. When comparing our results
to the outcome of similar investigations, one finds that quite different
theoretical approaches to the reaction dynamics finally result in the
similar dilepton spectra. For example the fireball approach has a
significantly different evolution of $T$, $\mu_{\mathrm{B}}$ and
$\mu_{\pi}$ but the resulting dilepton excess spectrum shows no large
deviations from our result (compare \cite{vanHees:2006ng}).  Furthermore
the relative strengths of the various contributions vary from the
coarse-graining results. It therefore becomes clear that the
time-integral character of the dilepton spectrum is not only a challenge
for theory but also makes it difficult to discriminate between the
models. Nevertheless, there is general agreement in line with our
findings that hadronic modifications of the $\rho$ spectral function are
the main source of the low-mass dilepton excess as observed at SPS
energies. The comparison of the two spectral functions indicates that
for a full description of the medium modifications several hadronic
interactions have to be taken into account, as in the hadronic many-body
calculation.
\section*{Acknowledgments} 
The authors acknowledge R. Rapp for providing the spectral function and
for many fruitful discussions. This work was supported by BMBF,
HICforFAIR and the H-QM.
\bibliography{Bibliothek}
\end{document}